\documentclass[fleqn,usenatbib]{mnras}

\usepackage{newtxtext,newtxmath}

\usepackage[T1]{fontenc}
\usepackage{ae,aecompl}

\usepackage{graphicx}	
\usepackage{amsmath}	
\usepackage{amssymb}	
\usepackage{booktabs}
\graphicspath{{images/}}

\newcommand{\Msun}           {\,{\rm M}_\odot}

\title[How Unusual is the Milky Way?]{How Unusual is the Milky Way's Assembly History?}

\author[T. A. Evans et al.]{
Tilly A. Evans,\thanks{E-mail: tilly.evans@durham.ac.uk}
Azadeh Fattahi,
Alis J. Deason
and Carlos S. Frenk
\\
Institute for Computational Cosmology, Department of Physics, Durham University, South Road, Durham, DH1 3LE, UK
}

\date{Accepted XXX. Received YYY; in original form ZZZ}

\pubyear{2020}

\begin{document}
\label{firstpage}
\pagerange{\pageref{firstpage}--\pageref{lastpage}}
\maketitle

\begin{abstract}
  In the $\Lambda$CDM model of structure formation galactic haloes
  build up by accretion of mass and mergers of smaller subunits.  The
  most recent massive merger event experienced by the Milky Way (MW)
  halo was the accretion of the Large Magellanic Cloud (LMC; which has
  a stellar mass of $\sim \; 10^9M_\odot$). Recent analyses of
  galactic stellar data from the \textit{Gaia} satellite have
  uncovered an earlier massive accretion event, the
  Gaia-Enceladus-Sausage (GES), which merged with the MW around 10 Gyr
  ago. Here, we use the EAGLE cosmological hydrodynamics simulation to
  study properties of simulated MW-mass haloes constrained to have
  accretion histories similar to that of the MW, specifically the recent accretion of an ``LMC'' galaxy and a
  ``GES'' merger, with a
  quiescent period between the GES merger and the infall of the LMC
  (the ``GES \& LMC'' class). We find that $\sim 16$ per cent of MW-mass haloes
  have an LMC; $\sim 5$ per cent have a GES event and no further merger with
  an equally massive object since $z=1$; and only $0.65$ per cent belong to
  the LMC \& GES category. The progenitors of the MWs in this last
  category are much less massive than average at early times but
  eventually catch up with the mean. The LMC \& GES class of galaxies
  naturally end up in the ``blue cloud'' in the colour-magnitude
  diagram at $z=0$, tend to have a disc morphology and have a larger
  than average number of satellite galaxies.

\end{abstract}




\section{Introduction}

The Milky Way (MW) is sometimes regarded as a template for studies of
the structure and evolution of $\sim L^\star$ spiral galaxies. Yet,
the more we find out about the MW, the more we recognise that it is
anything but typical; in fact, several of its properties are
distinctly atypical. For example, the MW hosts a very massive
satellite, the Large Magellanic Cloud (LMC), which has approximately
10\% of the total MW halo mass
\citep[e.g.][]{benson2002,penarrubia_timing_2016,shao_evolution_2018}. Studies using large samples of observed local
galaxies, combined with cosmological simulations, find that only
$\sim 10\%$ of MW-mass galaxies host a satellite as massive as this
\citep[e.g.][]{liu_how_2011,busha_mass_2011,boylan-kolchin_too_2011,tollerud_large_2011}. Not only is the presence of
a massive satellite rare, but the scale of the damage the LMC is
inflicting on the Galaxy has recently began to be recognised. This
includes perturbing the barycentre of the MW \citep{gomez_and_2015},
disturbing the Galactic disc \citep{laporte_response_2018} and
inducing a large-scale gravitational ``wake'' in the stellar and dark
matter haloes \citep{garavito-camargo_hunting_2019}. Clearly the
accretion of the LMC is a significant, transformative event in our
Galaxy's history.
 
In addition to hosting the LMC, the MW has other peculiarities.  Its
central supermassive black hole has an abnormally small mass compared
to other galaxies of similar stellar mass \citep[e.g.][]{savorgnan_supermassive_2016}; its satellite system has a strange planar alignment
perpendicular to the MW disc
\citep[e.g.][]{lynden-bell_dwarf_1976,libeskind_distribution_2005,metz_orbital_2008}; and
the Galactic stellar halo may be unusually low-mass and metal-poor
(e.g. \citealt{bell_galaxies_2017,harmsen_diverse_2017}, but see
\citealt{conroy_resolving_2019,deason_total_2019}). Some of these
seemingly atypical qualities could be explained by the paucity of
mergers experienced by our Galaxy. For example, it is expected that
our forthcoming merger with the LMC (in a couple of gigayears time)
may return the MW back to ``normality''
\citep{cautun_aftermath_2019}. In view of all these peculiarities, a
natural question to ask is how similar or different is the MW assembly
history to that of other galaxies of similar mass?

A fundamental prediction of the $\Lambda$CDM model is that MW-mass
galaxies grow by accretion and mergers with smaller galaxies.
Simulations show that large stellar haloes form as a result of these
accretion events \citep{bullock_tracing_2005,
  abadi_stars_2006,font_phase-space_2006, cooper_galactic_2010}.
Dynamical timescales in the stellar halo are long, so the
phase-space distribution of halo stars can retain some memory of the
past accretion events. Moreover, the chemistry of the debris of merger
events reflects that of the progenitor galaxies: more massive dwarfs
are more metal-rich than lower mass dwarfs, and have distinct
sequences in chemical abundance space
\citep[e.g.][]{tolstoy_star-formation_2009}. Hence, by analysing the
stellar phase-space and chemistry properties, it may be possible to
identify different accretion events. The link between the
chemodynamics of halo stars and the Galaxy's assembly history can be
traced back to the early work by \cite{eggen_evidence_1962}. More
recently, our view of the Galaxy has been transformed by the
availability of 6D phase-space information for large numbers of halo
stars provided by large astrometric, photometric and spectroscopic
surveys.

The {\it Gaia} mission, in particular, is providing new detailed
insights into the assembly history of the MW. The first and second
data releases led to the discovery of an ancient merger event,
discovered independently by two teams who called it ``Gaia Enceladus''
\citep{helmi_merger_2018} and ``Gaia Sausage''
\citep{belokurov_co-formation_2018} respectively; here we will refer
to both jointly as the Gaia-Enceladus Sausage (GES)\footnote{Note,
  however, that it is still debated whether or not these two
  discoveries are describing exactly the same event
  \citep[e.g.][]{evans_early_2020,elias_cosmological_2020}}.

\citet{helmi_merger_2018} analysed the kinematics, chemistry and
positions of stars in the MW's thick disc and stellar halo, and found
that high-energy stars on retrograde orbits are also linked by their
chemical compositions. They concluded that the inner halo is dominated
by stars coming from a single object accreted around 8-11 Gyr
ago. \citet{belokurov_co-formation_2018} used kinematics and chemistry
of stars from SDSS and \textit{Gaia} DR1 to show that, at higher
metallicity ([Fe/H]>1.7), the orbits of the halo stars are very
radially biased (with velocity anisotropy, $\beta \sim 0.9$). These
authors argue that the extreme radial orbits in the inner stellar halo
cannot have been caused by steady accretion of low-mass dwarf
galaxies, but instead must have come from a single merger event with a
massive satellite some 8-11 Gyr ago. This proposal agrees with the
idea of orbit radialization put forward by
\cite{amorisco_contributions_2017}. The connection relation between
highly radial orbits and massive merger events has been confirmed with 
cosmological simulations
\citep[e.g.][]{fattahi_origin_2019,mackereth_origin_2019}.

In addition to halo stars, globular clusters can be used to identify
accreted dwarf galaxies since, by virtue of their high stellar mass
density, they are able to survive tidal disruption long after a dwarf
galaxy that brought them into the MW has been destroyed
\citep[e.g.][]{kruijssen2009,penarrubia_tidal_2009}. Recent work using
\textit{Gaia} DR2 shows that the Galactic globular cluster population
also points to a GES merger event \citep{myeong_sausage_2018,
  pfeffer_predicting_2020}. In particular, \citet{myeong_sausage_2018}
show that at least eight Galactic globular clusters are likely to be
associated with the GES. However, in addition to the GES, the globular
cluster population has possibly revealed two additional merger events:
Sequoia and Kraken \citep{myeong_evidence_2019,
  kruijssen_kraken_2020}. Sequoia is thought to have merged with the
MW around the same time as the GES, but the progenitor had a much
lower mass \citep{myeong_evidence_2019}. Kraken is thought to be much
older (accretion at $z > 2$), but may have an even higher mass ratio
relative to the MW than the GES merger 
\citep{pfeffer_predicting_2020}. The existence of this latter event is
still under debate, and it may be more difficult to identify in the halo
stars as its stellar debris likely occupies the inner few kiloparsecs
of the Galaxy.

In view of the recent advances in our knowledge of the assembly
history of the MW, the aim of this paper is to characterise MW-like
galaxies in cosmological simulations that have similar past accretion
events to our own galaxy. Specifically, we use the cosmological
hydrodynamics EAGLE simulation
\citep{schaye_eagle_2015,crain_eagle_2015} to identify MW-mass systems
that underwent mergers analogous to \textit{both} the LMC and GES
events.  Previous studies have considered MW-like galaxies with either
an LMC \textit{or} a GES event
\citep[e.g.][]{bignone_gaia-enceladus_2019, cautun_aftermath_2019,
  elias_cosmological_2020}, but not both as seems appropriate for the
actual MW. The ``classical'' view of the LMC
\citep[e.g.][]{cautun_aftermath_2019} allows for a variety of previous
mergers but does not include the paucity of massive merger events
experienced by the MW between the recent infall of the LMC and the
ancient merger of the GES. This paper aims to explore differences
amongst galaxies that experience late (LMC), early (GES) merger
events, or both. In particular, we aim to establish how unusual it is
for a MW-mass galaxy to have both these events and a dearth of massive
mergers in between.

The paper is arranged as follows. In Section~\ref{sec: simulations} we
describe the simulations used for our analysis and our sample
selection. In Section~\ref{sec: results} we present our results; we
discuss and summarise our conclusions in Section~\ref{sec: disc and
  conc}.

\section{Simulations}\label{sec: simulations}
We now give a brief overview of the EAGLE simulations.  We also
describe the selection criteria for our MW-mass galaxies, and the
reasoning behind each choice.

\subsection{EAGLE}

We aim to identify MW analogues in the EAGLE simulation
\citep{schaye_eagle_2015,crain_eagle_2015,mcalpine_eagle_2016}, which
have either an LMC-mass satellite, an ancient merger similar to the
GES, or both an LMC and a GES event.  The EAGLE project is a suite of
cosmological hydrodynamical simulations that follow the formation and
evolution of galaxies, tracking the gas, stars and dark matter
throughout cosmic history. The simulations use a modified version of
the Tree-PM SPH code {\sc p-gadget3}, which is based on the publicly
available code {\sc gadget-2} \citep{springel_cosmological_2005}. The
hydrodynamics solver uses a pressure-entropy formalism \citep[see][for
details]{schaller_eagle_2015}. The subgrid galaxy formation model
includes homogeneous photoionising background radiation,
metallicity-dependent star formation and cooling, stellar evolution
and supernovae feedback, seeding and growth of supermassive
black holes, and AGN feedback \citep[see][for a full description of the
model]{schaye_eagle_2015}. The galaxy formation model was calibrated
to reproduce the stellar mass function of galaxies at $z=0.1$ and
realistic galaxy sizes, down to a stellar mass of $\sim 10^8
\Msun$. The model has been shown to produce galaxies with realistic
mass profiles and rotation curves \citep{schaller_baryon_2015}.

We analyse the main EAGLE simulation (Ref0100N1504) which follows the
evolution of a periodic cubic volume of (100~Mpc)$^3$. The mass
resolution is $9.6\times10^6 \Msun$ and $1.81\times10^6\Msun$ for dark
matter and gas particles, respectively. The Friends-of-Friends (FoF)
algorithm was used to identify dark matter haloes
\citep{davis_evolution_1985}, and the {\sc subfind} algorithm was used
to identify self-bound structures and substructures within the FoF
groups \citep{springel_cosmological_2005}. The cosmological parameters
adopted for the simulations are based on
\citet{planck_collaboration_planck_2014}: $\Omega_m=0.307$,
$\Omega_\lambda=0.693$, $\Omega_{bar}=0.048$,
$H_0=67.77\,{\rm km\,s^{-1} Mpc^{-1}}$, $\sigma_8=0.8288$.

The large volume of the main EAGLE simulation provides a large number
of MW-mass galaxies which have a variety of merger histories.  The
stellar mass, M$_{*}$, for each galaxy was calculated by summing the
mass of the star particles bound to each galaxy within 30~kpc of the
galaxy's centre. The halo mass and radius, $M_{200}$ and $R_{200}$
respectively, are defined as those of the sphere with average enclosed
density equal to 200 times the critical density of the universe. We
define satellites as those subhaloes within $R_{200}$ of their
host. Infall time or accretion time is defined as the time when a
galaxy crosses the $R_{200}$ radius of the host. In practice, we take
the snapshot immediately before the crossing. Finally, merger
time is defined as the last snapshot at which an accreted galaxy was 
identified by {\sc subfind}, before it merges with the host or is
destroyed. 

\begin{figure}
    \centering
    \includegraphics[width=\columnwidth]{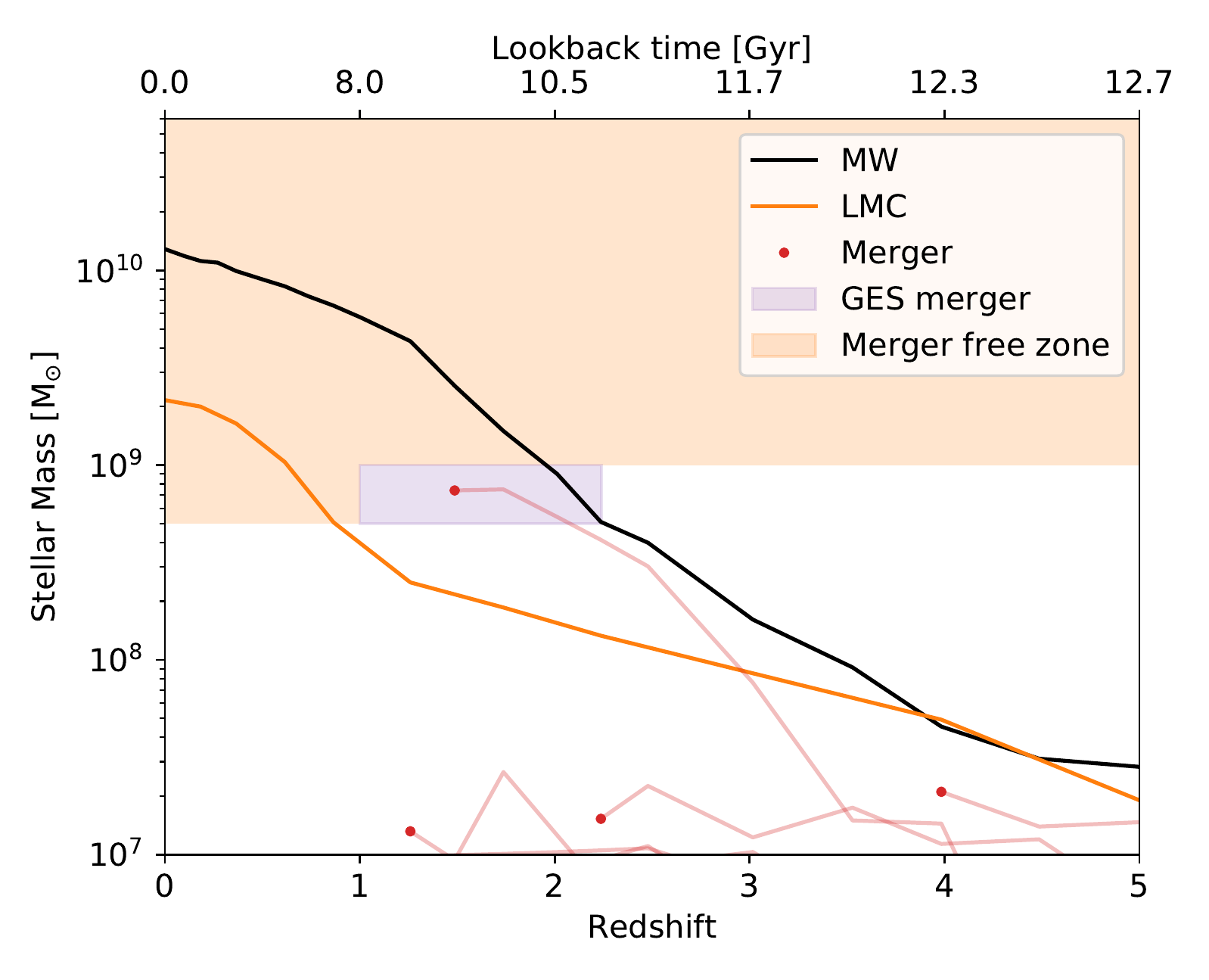}
    \caption{An example of mass growth and merger history of a MW-mass
      galaxy in EAGLE. The black and orange solid lines show the
      stellar mass of the main MW progenitor and its LMC satellite
      from $z=5$ to $z=0$. Light red lines correspond to all dwarf
      galaxies (above stellar mass of $10^7\Msun$) that merged with
      the MW halo. Red dots mark the merger time (i.e. the latest
      snapshot where the progenitors were identified). The orange
      shaded region illustrates the area which, according to our
      criteria (sec. \ref{sec: criteria}), should be merger free. The
      purple shaded region illustrates the criteria for when a GES
      merger should occur.}
    \label{fig: merger_history}
\end{figure}

\subsection{Sample selection}\label{sec: criteria}
We first identify all galaxies with a present-day dark matter halo in
the mass range $(0.7-2)\times 10^{12} $M$_{\odot}$ \citep[see][and
references therein]{callingham_mass_2019}. This sample comprises $N=1078$ haloes. We then impose further restrictions based on assembly
history.

The LMC is a relatively massive satellite,
$M_{*}=1.5\times10^9 \Msun$, with a small Galactocentric distance,
$d_{GC}=50$ kpc, and moves with a large tangential speed
\citep{mcconnachie_observed_2012,kallivayalil_third-epoch_2013}. The
Galactocentric distance and velocity of the LMC indicate that it is
close to the pericentre of its orbit. We do not impose any constraint
on the orbit of LMC-mass satellites and only require that the stellar
mass of the satellite galaxy should lie in the range
$(1-4)\times 10^9 $M$_{\odot}$, and the satellite be located within
$R_{200}$ of the host halo at $z=0$.  We additionally require that no
other satellite more massive than the LMC exists within the $R_{200}$
radius of the MW-mass hosts at $z=0$. We do not impose a constraint
on the infall time of the LMC-mass satellite. However, since these
massive satellites survive to $z=0$, they typically infall at late
times (see Table~\ref{tab: analogue info}).

The details of the GES-like event are more uncertain since it is a
fairly new discovery. However, the stellar mass is likely to be in the
range $(0.5-1) \times 10^9 $M$_{\odot}$ and it is thought to have
merged with the MW between 8 Gyr and 11 Gyr ago
\citep{belokurov_co-formation_2018,fattahi_origin_2019,mackereth_origin_2019}\footnote{\citet{fattahi_origin_2019}
  find a slightly higher stellar mass for the GES progenitors in the
  Auriga simulations. However, the stellar masses in Auriga subhaloes
  are slightly overestimated.}. We impose no orbital constraint on the
GES event.  Similarly to the LMC constraint, we require that there was
{\it only one} GES-mass accretion event in this time period (with no
merger with a galaxy more massive than the GES). Finally, we impose
the condition that there should be no massive accretion events with
progenitor stellar mass $M_*>0.5\times10^9\Msun$, in the interval
between the GES merger event and the infall of the LMC. Our
constraints do, however, allow lower mass accretion events such as Sgr
and Sequoia to occur between the merger with the GES and the infall of
LMC. We do not consider the accretion of a Kraken-like event since it
is suspected that this happened before the GES event. A brief
summary of the selection criteria are as follows:
\begin{enumerate}
\item  MW analogue has M$_{200} = (0.7-2)\times 10^{12} $M$_{\odot}$.
\item  LMC exists at $z=0$ within R$_{200}$, with stellar mass M$_{*} = (1-4) \times 10^9 $M$_{\odot}$.
\item  GES merger event of mass M$_{*} = (0.5-1) \times 10^9   $M$_{\odot}$ occurs at $t =8-11$ Gyr. 
\item A `merger-free zone' when there are no massive accretion
  (M$_*>0.5\times10^9\Msun$) events is required between the time of
  the merger of the GES and the infall of the LMC
\end{enumerate}

In order to apply these criteria, we used merger trees to follow the
assembly history of the MW-mass
galaxies. Fig.~\ref{fig: merger_history} shows an example of MW-mass
galaxy in EAGLE, and helps to visualise the constraints imposed by the
different selection criteria. The black and orange lines show the main
branch of the simulated MW galaxy and its LMC satellite
respectively. The red dots represent dwarf galaxies (above stellar
mass of $10^7\Msun$) which have merged onto the MW main branch, the
light red lines show the main branch of the merging dwarfs up until
they have completely merged with the MW. The orange and purple shaded
regions show the mass and redshift zones for the merger-free area, and
the area in which a GES merger should occur.

Throughout this paper, the properties of MW-mass galaxies are compared
in the ``categories'' described below. These groupings were chosen to
clarify how having either, or both of the LMC and GES would have influenced
the Galaxy's evolution. These categories are: 
\begin{enumerate}
\item All MW-mass galaxies (MW-all).
\item MW-mass galaxies which have an LMC within R$_{200}$ at $z=0$
  with no merger history constraint\footnote{This is the ``classical''
    MW with LMC studied in many previous papers.} (LMC-all). 
\item MW-mass galaxies which have an LMC within R$_{200}$ at
  $z=0$, but did not experience a GES merger event or accrete any
  other massive  dwarfs according to our constraints (LMC-o). 
\item MW-mass galaxies which have a merger event similar to the GES,
  but do not have an LMC satellite at $z=0$ and did not accrete any
  other massive dwarf according to our constraints (GES-o).
\item MW-mass galaxies which have both a LMC, a GES event, and a
  `merger free zone' during which no object more massive than the LMC
  or the GES is accreted. This final group (LMC \& GES) most closely describes the
  true MW galaxy\footnote{\citet{kruijssen_kraken_2020} found a lower
    mass for the GES than used in this paper. However, their findings for
    the MW assembly history are consistent with the LMC-o sample of
    galaxies presented here.}.
\end{enumerate}

\begin{table*}
\centering
\caption{Details of the MW-mass galaxies included in our sample}
\renewcommand\arraystretch{1.5}
\begin{tabular}{c c c c c c}
\toprule
Group       & Number of galaxies & Percentage & Median $M_{*}$  & Median $M_{200}$ & $M_{*}/M_{200}$ \\
 & & & ($\times 10^{10}$ M$_{\sun}$) & ($\times 10^{12}$ M$_{\sun}$) &($\times 10^{-3}$) \\
 \midrule
MW-all   &  1078             &   -            & 2.03       &      1.04        &19.5\\ 
LMC-all     &   169             &   15.7\%       & 1.78       &      1.12       & 15.9\\ 
LMC-o        &   40              &   3.7\%        & 1.30       &      0.87       & 14.9\\
GES-o         &   54             &   5.0\%       & 1.74       &      0.92        &18.9\\
LMC \& GES        &   7              &   0.65\%        & 1.44       &      0.80    &   18.0 \\ \bottomrule

\end{tabular}
\label{tab: percentages}
\end{table*}

The number of galaxies that meet these criteria is presented in
Table~\ref{tab: percentages}, along with the percentage of MW galaxies
that belong in each category and their median stellar and halo
masses. Our values for the fraction of MW-mass haloes in the LMC-o or
GES-o categories are significantly higher than those stated in
\citet{bignone_gaia-enceladus_2019} and
\citet{cautun_aftermath_2019}. \citet{cautun_aftermath_2019} found
only eight MW analogues in the EAGLE simulation which had an LMC-mass
satellite; however, that work applied more restrictive constraints to
the sample. For example, the mass of the cold gas content and the
black hole mass were also considered in the constraints.
\citet{bignone_gaia-enceladus_2019} found only one MW galaxy with a
GES type merger event in the EAGLE simulation; however, that work also
imposed constraints on the current star formation and the disc of the
MW, and, importantly, required that the stellar debris from the GES
event be highly anisotropic.  

Our constraints are deliberately imposed to depend only on the mass
and time of significant accretion events in the MW's history. Anything
more restrictive would result in a very small sample size. 
The
fraction of MW haloes with `classical' LMCs (LMC-all) is slightly
higher than the observed value ($\sim 10\%$, see e.g. \citealt{liu_how_2011, tollerud_large_2011}) but,
again, our constraints are less restrictive (e.g. no constraint on present-day position, or orbit of the LMC).
  The final value for the fraction of MW
  galaxies with both an LMC and a GES merger event, and nothing
  significant in between, is only 0.65\% of all MWs in EAGLE -- this
  is already an indicator that our Galaxy's assembly history is very
  rare.

\begin{figure*}
    \centering
    \includegraphics[width=\textwidth]{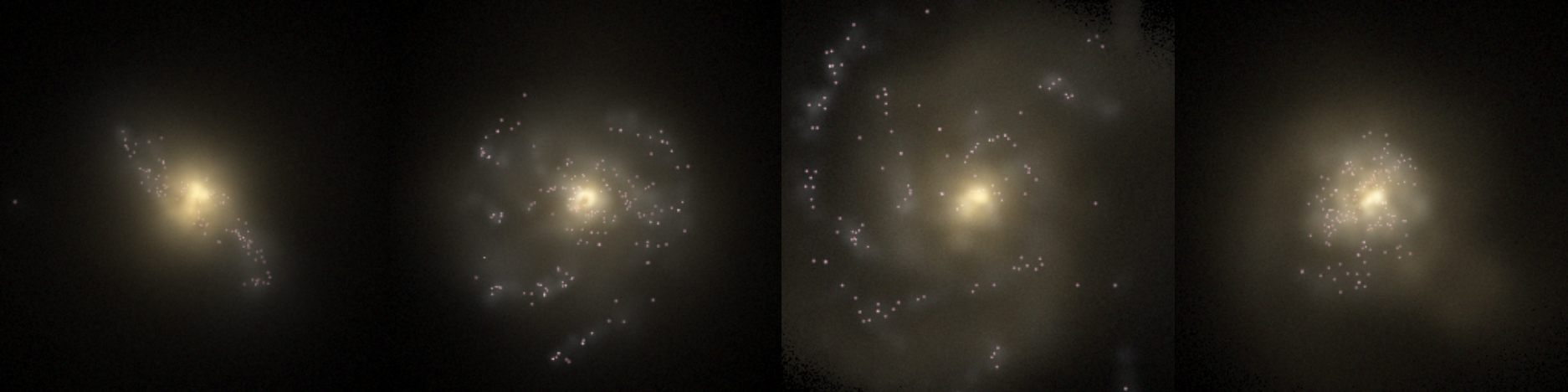}
    \caption{Composite face-on $gri$ images of four representative
      MW-mass galaxies in our final sample (LMC \& GES) from the EAGLE
      reference simulation. Each panel is $60 \times 60$ pkpc
      wide. Details of the visualisation can be found in
      \citet{trayford_optical_2017}. Left to right, the EAGLE halo
      ID's are 9293658, 9372228, 9798319 and 10058549.}
    \label{fig: mock gri}
\end{figure*}

  Four representative galaxies from our final LMC \& GES sample are
  illustrated in Fig.~\ref{fig: mock gri} and Fig.~\ref{fig: dm
    galaxies}. The former shows mock $gri$ face-on images of the
  central galaxies at $z=0$, where face-on has been defined according
  to the stellar angular momentum axis
  \citep{trayford_optical_2017}\footnote{Note that
    \cite{trayford_optical_2017} states that the galactic plane is not
    always easily defined, which could explain the irregularity of the
    leftmost image in Fig.~\ref{fig: mock gri}}. The images have been
  retrieved from the EAGLE public
  database\footnote{\url{http://virgodb.dur.ac.uk:8080/Eagle/}}. They
  have been produced by post-processed ray-tracing using a version of
  the code {\sc SKIRT} \citep{camps_skirt_2015}; please see
  \citet{trayford_optical_2017} for more details.  Fig.~\ref{fig: dm
    galaxies} shows 2D projections of the dark matter particles in a
  $(500 \mathrm{kpc})^3$ region around the four haloes. The outer and
  inner circles mark the $R_{200}$ and $0.5\times R_{200}$
  boundaries. The LMC-mass satellite is highlighted in each halo
  with a red circle.

\begin{figure*}
    \centering
    \includegraphics[width=\textwidth]{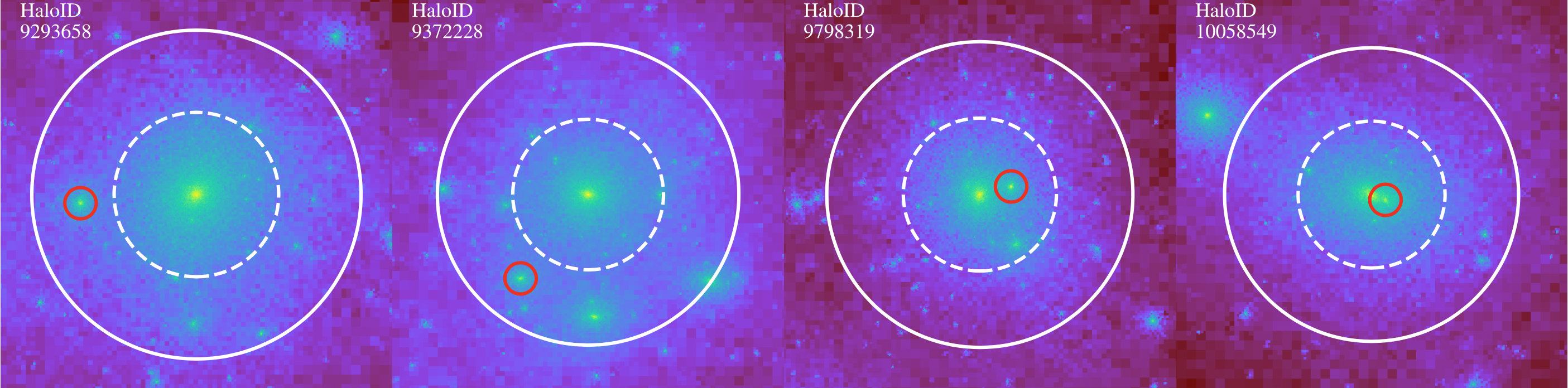}
    \caption{Distribution of dark matter particles for the
      corresponding galaxies shown in Fig. \ref{fig: mock gri}. The
      solid and dashed white circles represent the host R$_{200}$
      and $0.5 \times$R$_{200}$ respectively. The smaller red circle
      shows the location of the LMC satellite. The EAGLE halo ID is
      shown in the top left-hand corner of each halo.} 
    \label{fig: dm galaxies}
\end{figure*}

\section{Results}\label{sec: results}
\begin{figure*}
    \centering
    \includegraphics[width=\textwidth]{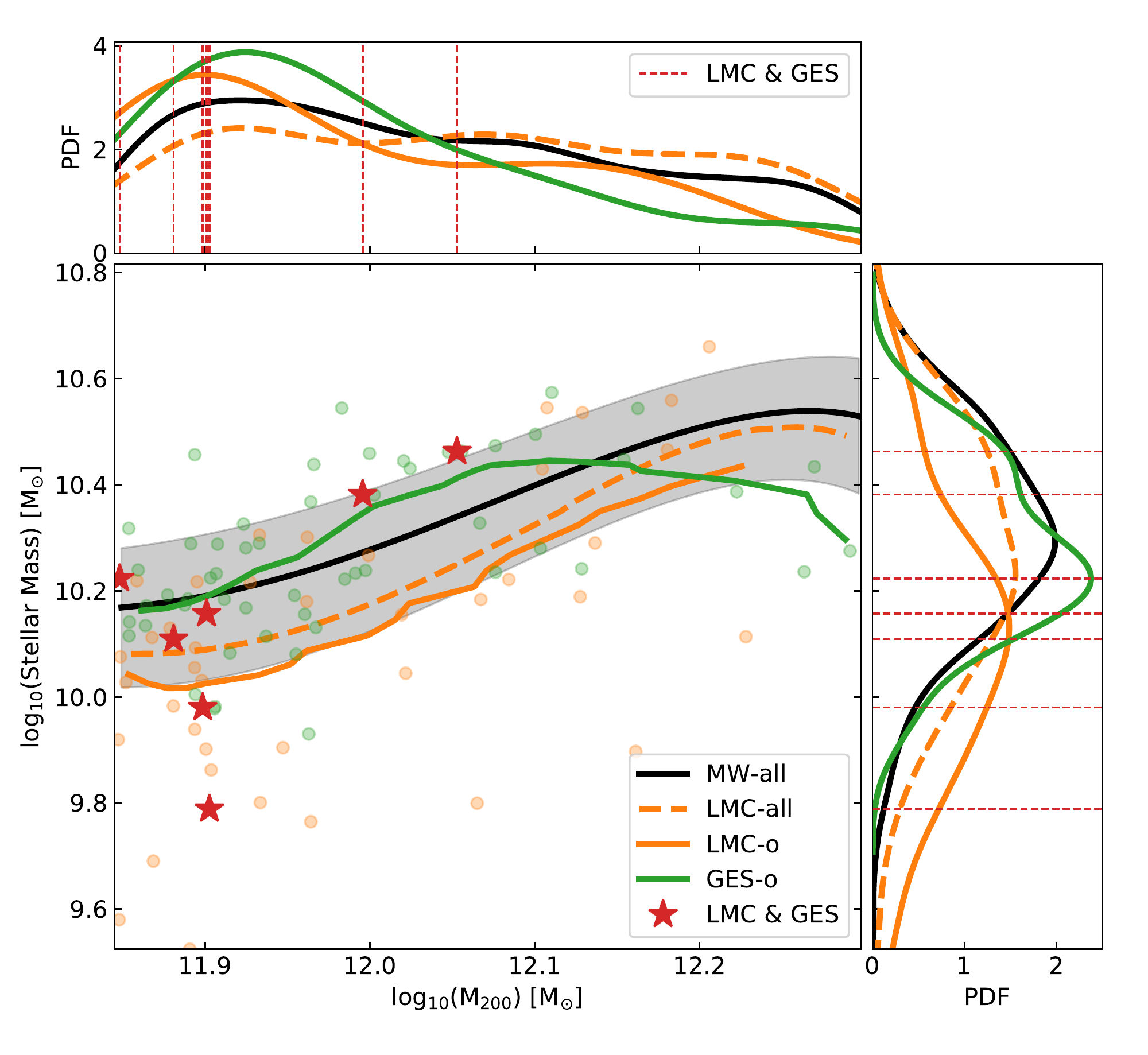}
    \caption{\textit{Central:} stellar mass-halo mass relation at
      $z=0$ for our different samples of MW-mass haloes (see
      Sec.~\ref{sec: criteria} for full details), as indicated in the
      legend.  The lines represent the median values of the stellar
      mass in halo mass intervals. For the `MW-all' sample we show the
      median absolute deviation (MAD) around the median as a shaded
      region, whereas for the other samples we show individual galaxies
      with the corresponding colour.  The medians and the MAD were
      smoothed using a Savitzky-Golay filter \citep{savitzky_smoothing_1964}.  The red stars
      correspond to the individual galaxies in the final sample of MWs with LMC \&
      GES. \textit{Top:} $M_{200}$ distribution of various MW-mass
      samples, smoothed with a KDE kernel. Vertical lines indicate
      individual galaxies in our LMC \& GES sample. \textit{Right:}
      similar to the top panel but showing the stellar mass distribution of
      MW-mass haloes.}
    \label{fig: star_halo hists}
\end{figure*}

\begin{table*}
\centering
\caption{The final sample of 7 MW analogues with 
  LMC \& GES. The columns give: (a)~EAGLE halo ID, corresponding to the
  halo ID in the top left of Fig.~\ref{fig: dm galaxies};
  (b)~$M_{200}$ of the MW halo at $z=0$;  (c)~the maximum stellar mass
  recorded for the destroyed GES; (d)~the redshift at which the GES 
  merged; (e)~maximum stellar mass of the LMC analogue; (f)~redshift
  at which the LMC analogue crossed the $R_{200}$ radius of the MW; and
  (g)~the number of satellites within $R_{200}$ of the MW analogues
  with stellar mass, $M_{*}>10^6\Msun$.}  
\renewcommand\arraystretch{1.5}
\begin{tabular}{c c c c c c c c}
\toprule
 EAGLE & MW M$_{200}$       & GES $M_{*}$    &   GES merger  & LMC $M_{*, max}$ &LMC infall   & Number of\\ 
 Halo ID &($\times 10^{12}$ M$_{\odot}$)  &  ($\times 10^{8}$ M$_{\odot}$) & redshift &  ($\times 10^{9}$ M$_{\odot}$) &redshift & satellites\\
(a) & (b) & (c) & (d) & (e) & (f) & (g)  \\
 \midrule
8806615  &1.13 & 5.2 & 2.01 &2.16&0.37&6\\ 
9293658  &0.99 & 2.9 & 1.00 &1.29&0.27&8\\ 
9372228  &0.76 & 7.4 & 1.49 &2.16&0.27&9\\
9626514  &0.80 & 6.2 & 1.00 &1.26&0.1&13\\ 
9798319  &0.80 & 9.0 & 1.74 &1.27&0.74&12\\
9968042  &0.79 & 7.8 & 1.00 &1.46&0.37&6\\
10058549  &0.71 & 9.9 & 1.26 &3.34&0.62&10\\
\bottomrule

\end{tabular}
\label{tab: analogue info}
\end{table*}

We now present an overview of the properties of our various MW-mass
categories defined in the previous section and their assembly
histories. We start by examining the stellar and halo mass distributions
and the stellar mass -- halo mass relation.

The halo mass distributions of our different MW-mass categories are
displayed in the top panel of Fig.~\ref{fig: star_halo hists}. The
black line shows the distribution of halo masses for the MW-all
sample. The mass distributions are approximately linear, with a slight
bias towards lower mass haloes, as expected from the power law halo
mass function in $\Lambda$CDM in this mass range
\citep{jenkins_mass_2001}. The decrease in halo mass distributions at the
lowest mass bins are an artefact introduced by the KDE kernel used for
smoothing. The LMC-all sample (dashed orange line) follows the general
MW-mass halo trend closely, although it is flatter and less biased
towards lower masses. This classical LMC-all group is different from
the LMC-o group shown as the solid orange line, which has a more
prominent peak towards lower masses.  

Fig.~\ref{fig: star_halo hists} shows that the halo mass of the MW is
shifted towards lower values for both the LMC-o and GES-o samples (medians
of $0.87\times10^{12}\Msun$ and $0.92\times10^{12}\Msun$
respectively). This effect is further enhanced when the MW has both an
LMC and a GES. The seven LMC \& GES galaxies are represented
individually by vertical dashed red lines. Halo mass, by definition,
is all the mass within $R_{200}$ and therefore includes the halo mass
of the LMC-mass satellite.

The stellar mass-halo mass relationship is shown in the central panel
of Fig.~\ref{fig: star_halo hists}.  The LMC-all and LMC-o samples
appear to have a lower stellar mass compared to the total sample at
any given halo mass. We, however, note that there are biases in the
way halo and stellar masses are measured. The stellar mass includes
the stellar particles within 30~kpc of the centre of the MW; however,
the halo mass, $M_{200}$, includes all mass within the FoF group
(i.e. including the halo of the LMC which is generally more than
30~kpc away from the centre). Therefore, the difference is not a shift
down in stellar mass for MWs with an LMC but rather a shift to the
right in the stellar mass-halo mass plane.

The central panel of Fig.~\ref{fig: star_halo hists} suggests that the
GES-o category has a higher than average stellar mass for a given halo
mass in the range ${\rm log}_{10}(M_{200}) = 11.9-12.1$. However, the
stellar mass distribution shown in the right panel shows that there is
a small peak at slightly higher stellar mass for the GES-o sample than
the average MW peak, but the main peak for the GES-o sample is at a
slightly lower stellar mass than the average. This is because this
category is biased towards lower halo masses (see top panel), and thus
lower, on average, stellar masses. For low halo masses
(${\rm log}_{10}(M_{200})<12$), the LMC \& GES sample has lower than
average stellar mass, but for higher halo masses
(log$_{10}(M_{200})>12$), it has higher than average stellar mass. The former is due to the aforementioned reason for LMC samples; namely, the stellar mass is measured for the central galaxy but halo mass includes the mass of LMC.  

The right panel of Fig.~\ref{fig: star_halo hists} shows the
distribution of stellar masses for each category of galaxies. The
LMC-o sample has a peak at lower stellar masses than the MW-all
sample, reflecting the lower halo masses seen in the top panel. The
stellar masses of the LMC \& GES sample are distributed across the
entire range of stellar masses.

\begin{figure*}
    \centering
    \includegraphics[width=\textwidth]{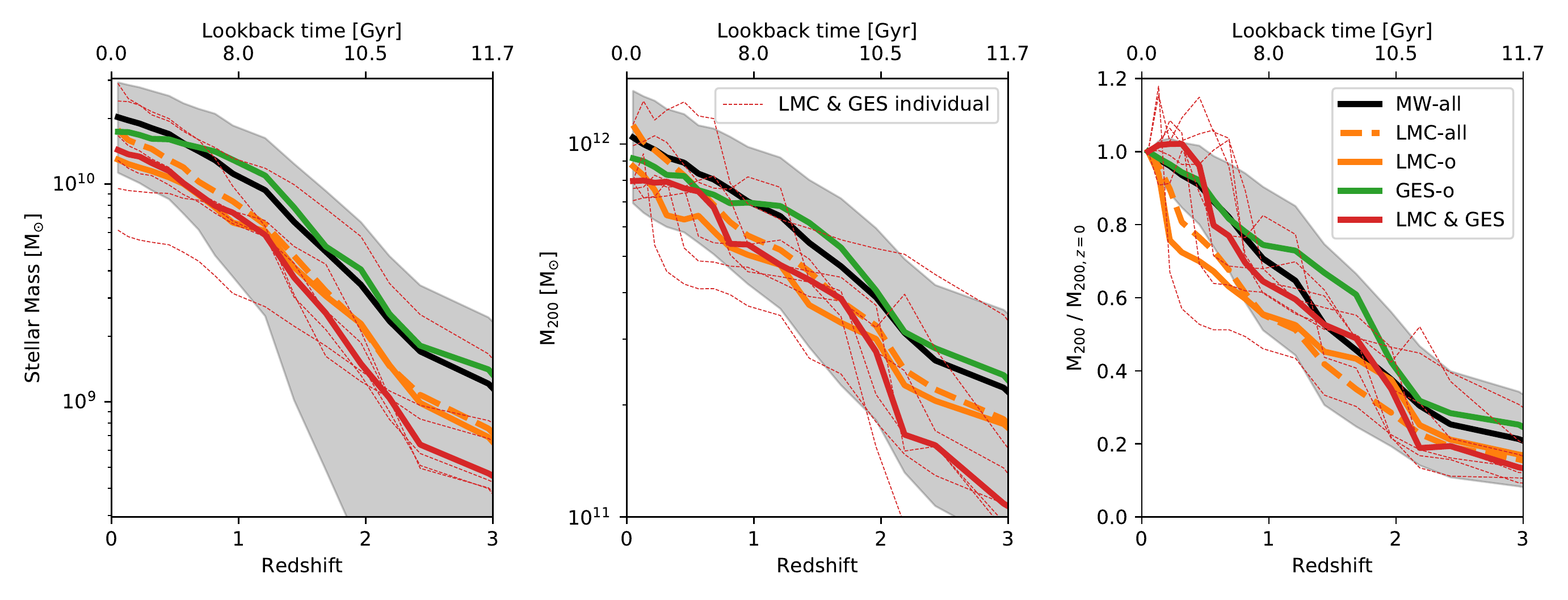}
    \caption{\textit{Left:} median stellar mass as a function of
      redshift for different MW-mass samples. The colour coding and
      line styles are similar to those in Fig.~\ref{fig: star_halo
        hists}. The shaded region represents the MAD around the median
      for the MW-all sample. The thin red dotted lines correspond to
      individual galaxies in the final LMC \& GES sample.
      \textit{Middle:} as the left panel, but for the halo mass,
      $M_{200}$, as a function of redshift. \textit{Right:} as the
      middle panel, but for the halo masses normalised to the $z=0$
      value for each halo, as a function of redshift. }
    \label{fig: m200 vs z}
\end{figure*}

\begin{figure*}
    \centering
    \includegraphics[width=\textwidth]{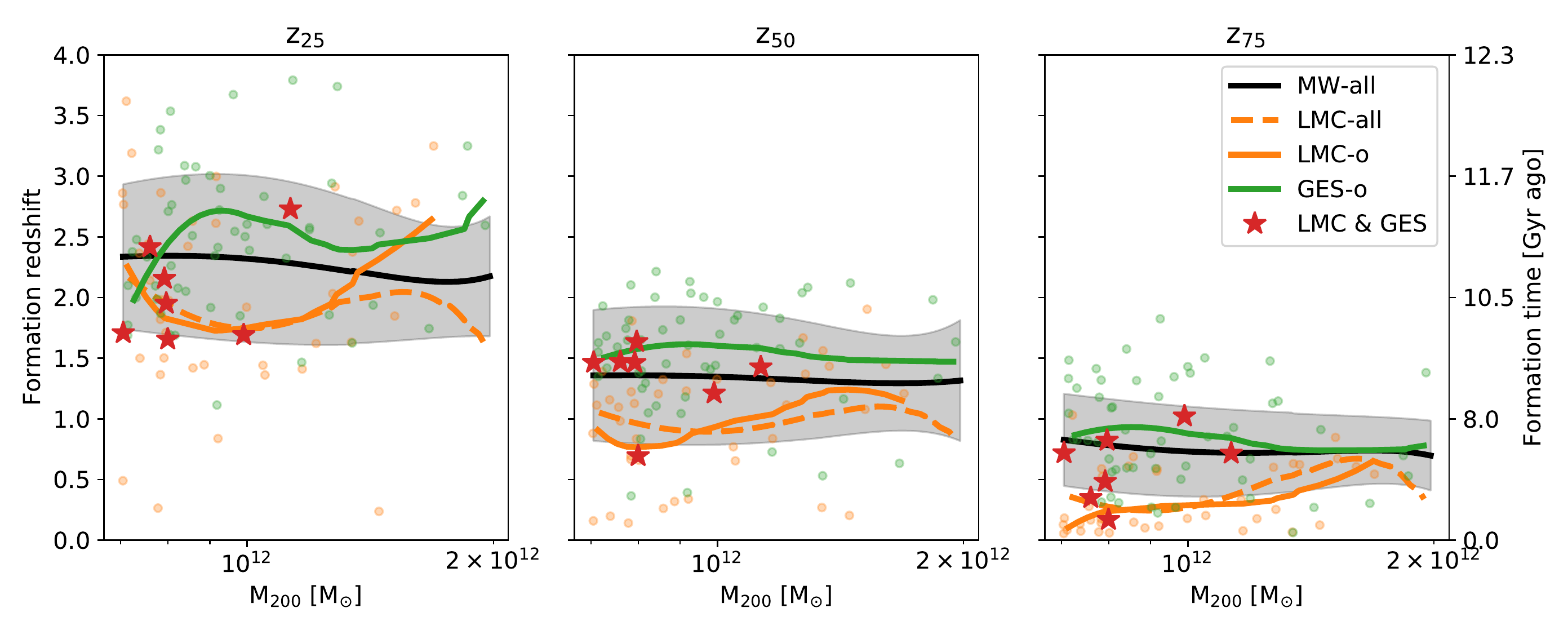}
    \caption{The formation redshifts of the MW-mass
      haloes. \textit{Left:} redshift when the halo has reached 25\%
      of its $z=0$ mass. \textit{Middle:} redshift when the halo has
      reached 50\% of its $z=0$ mass. \textit{Right:} redshift when
      the halo has reached 75\% of its $z=0$ mass. Colours, lines and
      symbols are similar to those of Fig.~\ref{fig: star_halo
        hists}. The black line represents the median value of the
      redshifts for the MW-all category, and the grey shaded region
      represents the MAD for that category. The median lines were
      smoothed using a Savitzky-Golay filter.}
    \label{fig: formation_times}
\end{figure*}

\subsection{Assembly history}

Fig.~\ref{fig: m200 vs z} shows the evolution of stellar mass (left
panel), $M_{200}$ (middle panel), and the normalised halo mass
($M_{200}$/$M_{200, z=0}$; right panel) for each category of MW-mass 
galaxies. The thick lines show the median stellar mass and halo mass 
at each redshift for each sample.
 
 The LMC-all category (dashed orange line) does not have any
 constraint on assembly history and is consistently lower in stellar
 and halo mass until the infall of its LMC-mass satellite. During
 infall, the LMC satellite gives rise to a sudden increase in halo
 mass, which leads to the LMC-all sample having the highest halo mass
 of all categories at $z=0$. The LMC-o sample is consistently low in
 both stellar and halo mass. As in the case of the LMC-all sample, the
 LMC-o galaxies also show an increase in halo mass as the LMC-mass
 satellite is accreted by its host. However, unlike the LMC-all
 sample, the final halo masses are not particularly high, and, in
 fact, are lower on average than the MW-all sample. This difference
 reflects the fact the the LMC-o sample is constrained to have no other
 significant merger (other than the LMC), which naturally biases 
 this sample to lower halo masses. 

 For the most part, the GES-o group follows the average MW but there
 is a clear bump in halo mass (central and right panels of
 Fig.~\ref{fig: m200 vs z}) that appears roughly at the redshift of
 the merger with the GES. The stellar and halo masses of this category
 tends to follow the average MW until after the GES merger event has
 ended. Preceding this event, the stellar and halo mass rise more
 slowly than the average MW sample since the MW-all sample experiences
 more continuous mergers than the GES-o, and therefore has a
 continuous influx of accreted mass.

 Initially, the LMC \& GES galaxies have stellar and halo masses that
 are lower than those of the LMC-o sample. These galaxies then follow
 the evolution of the LMC-o category until the merger with the GES at
 $z=1-2$, when it also shows a bump in halo mass similar to that in
 the GES-o category. Some of the individual LMC \& GES galaxies show a
 very sudden increase in halo mass (in some cases almost double in
 mass) because of the accretion of the LMC-mass satellite. The stellar
 mass growth is much smoother than the halo mass growth. There is no
 bump in the stellar mass corresponding to the bump in halo mass since
 not all stars go to the centre of the MW (i.e. within 30kpc, the
 aperture used to calculate the stellar mass). The combination of the
 two events, the merger with the GES and the infall of the LMC
 increase the halo mass of the MW above the threshold of our
 definition MW-mass haloes. However, our results suggest that the main
 progenitor of the MW would have been much smaller than those of the
 other categories of MW-mass galaxies.

 The difference in the assembly histories of our various MW categories
 is more apparent in the right-hand panel of Fig.~\ref{fig: m200 vs
   z}, which shows the normalised halo mass as a function of
 redshift. There is a clear bump in the GES-o sample (green line)
 around the redshift of the GES merger with the MW. There is also
 clearly a sudden increase in halo mass of the LMC-o sample (solid
 orange line) at infall time of the LMC-mass satellite. Some of the
 LMC \& GES individual galaxies (red dashed lines) go above
 $M_{200}/M_{200, z=0} = 1$, which is likely due to the infall 
 of the LMC-mass satellite, followed by tidal stripping.

 We use the normalised halo mass assembly to define halo formation
 redshifts, $z_{X}$, for each MW-mass galaxy as the time when the main
 progenitor branch of the merger tree reaches X percent of the
 present-day M$_{200}$. Fig.~\ref{fig: formation_times} shows
 $z_{25}$, $z_{50}$, and $z_{75}$ of our MW samples in the left,
 middle and right panels respectively.

 Due to the sudden increase in halo mass resulting from the infall of
 the LMC-mass satellite, the formation redshifts of the LMC-all and
 LMC-o categories are shifted to lower values across all three
 panels. The $z_{75}$ formation redshift is the most affected by this
 since this is roughly the infall time of the LMC-like satellites,
 particularly for the lower halo mass galaxies. 

The GES-o sample has a consistently earlier formation redshifts across
all three panels. The left panel suggests that the GES-o galaxies form
25\% of their mass earlier than the other samples even before the
merger with the GES. The middle panel shows that the $z_{50}$
formation redshift occurs at roughly the same redshift as the merger
with the GES, and the right panel shows that after the merger with the
GES, the MW-mass galaxies follow the MW-all sample more closely. 

Overall, Fig.~\ref{fig: formation_times} suggests that the LMC \& GES
galaxies tend to have later $z_{25}$ formation redshifts, similar to
the LMC-o formation redshifts. This is expected since in that time
interval the GES is only beginning to merge with the MW and so the
galaxy has experienced roughly the same amount of merging as the LMC-o
category. The middle panel shows that the $z_{50}$ formation redshifts
of the LMC \& GES follow more closely the GES-o line, which is likely
due to the merger of the GES occurring at roughly these
redshifts. Finally, the right panel shows that the LMC \& GES return
to follow the LMC-o lines for the $z_{75}$ formation redshifts, likely
due to the imminent infall of the LMC-like satellites. Figs.~\ref{fig:
  m200 vs z} and~\ref{fig: formation_times} show that the assembly
history of the MW intricately follows the details of these two (GES \&
LMC) accretion events, and, at most redshifts, looks very different to
the average MW-mass halo.

\begin{figure*}
    \centering
    \includegraphics[width=\textwidth]{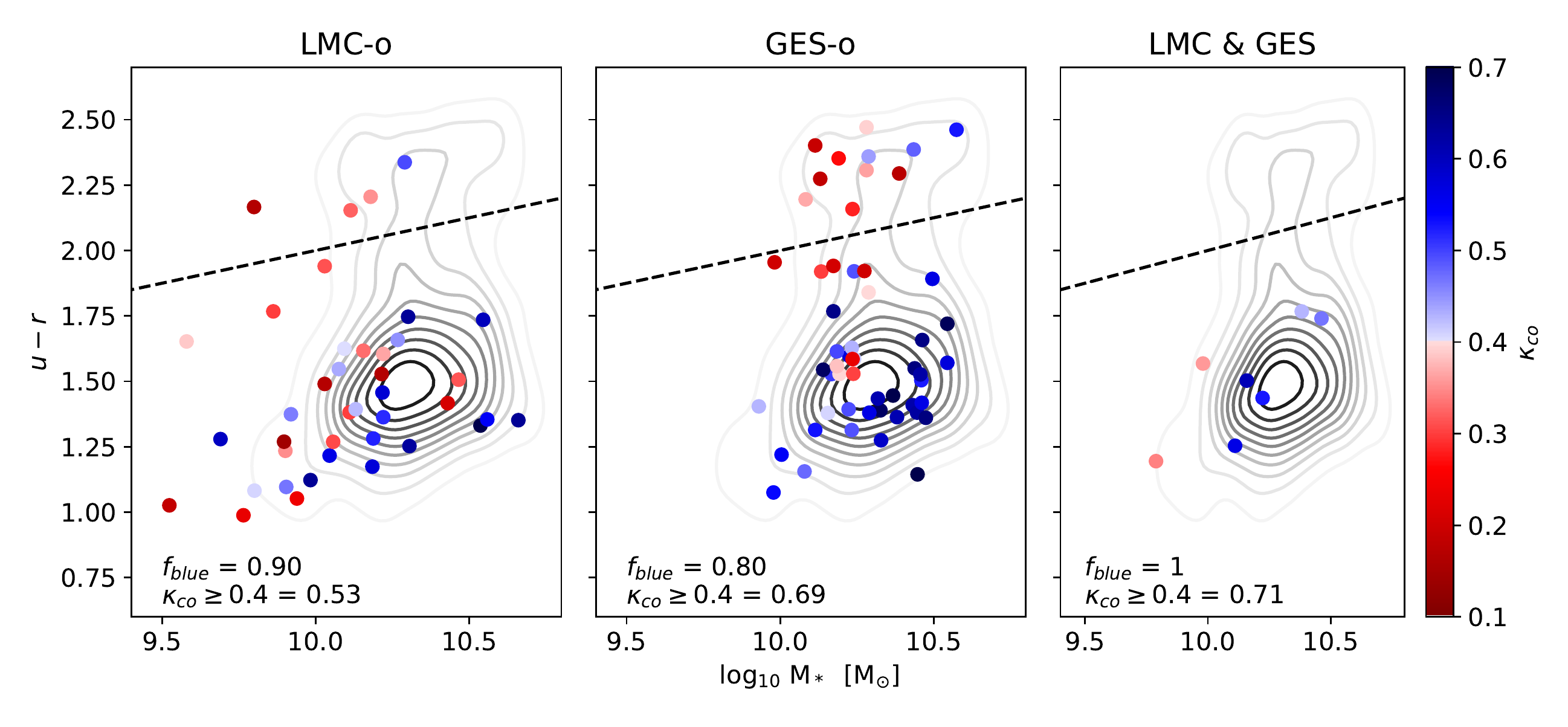}
    \caption{The $u-r$ colour {\em vs} stellar mass diagram for MW-mass
      galaxies. The panels correspond to the LMC-o (left), GES-o
      (middle) and LMC \& GES (right) samples. 
    The grey contours are repeated in all panels and show the
    colour-stellar mass distribution for the `MW-all' sample. Galaxies
    are represented by red or blue according to a cut 
    in their $\kappa_{co}$ values at $\kappa_{co}=0.4$, as shown by
    the colour bar. The dashed black line marks the separation of 
    the red sequence from the blue cloud. The legend gives the 
    fraction of blue galaxies (${\rm f}_{\rm blue}$) and the fraction
    of disc galaxies ($\kappa_{co} \geq 0.4$).}  
    \label{fig: colour_star}
\end{figure*}
\subsection{Galaxy colour and morphology}

Assembly histories are reflected in the morphology of galaxies. Our
selection of samples of MW-mass galaxies with different constraints
(Section~\ref{sec: criteria}) does not include any criteria for the MW
to be a star-forming spiral galaxy rather than a red, non-star forming
elliptical galaxy. In this section, we show how the assembly history 
affects the $z=0$ colours of our galaxies.

To characterize the MW-mass galaxies we consider their colour and
morphology.  The rest-frame absolute magnitude without dust
attenuation was used to estimate the colour; see
\citet{trayford_colours_2015} for details. We adopt the threshold
defined by \citet{schawinski_green_2014} to label galaxies as blue or
red (dashed line in Fig.~\ref{fig: colour_star}). The fraction of blue
galaxies for the MW-all sample is f$_{\rm blue} = 0.82$.

Fig.~\ref{fig: colour_star} shows the colour-stellar mass diagram for
our MW-mass haloes. The grey contours, repeated in all panels,
correspond to the distribution for the MW-all group which is clearly
bimodal showing a red sequence and a blue cloud. Different panels
correspond to the various categories of MW-mass galaxies, as labelled,
with the colour of each point indicating whether they are disc (blue)
or elliptical (red) galaxies. To characterise morphology we use the
stellar kinematics morphological indicator introduced by
\citet{sales_origin_2012} and calculated for EAGLE galaxies by
\citet{correa_relation_2017}. The fraction of disc galaxies (i.e. with
$\kappa_{co} \geq 0.4$) for the MW-all sample is 0.6.

The left panel shows that the LMC-o sample contains more blue galaxies
than average, with f$_{\rm blue}=0.9$. This is likely due to the late
formation redshift discussed previously, and the restriction that
there should be no significant mergers until the LMC infall. However,
fewer galaxies in the LMC-o sample are disc galaxies compared to the
MW-all sample. The middle panel shows that the GES-o sample contains a
similar fraction of blue galaxies as the MW-all sample but a much
higher fraction of disc galaxies. Since GES-o galaxies experience an
early merger event (and hence have an earlier formation redshift) they
are more likely to become redder with time. The right panel shows that
the LMC \& GES galaxies are all in the blue cloud. This is consistent
with the trend seen in the LMC-o sample that the later forming
galaxies are more likely to be bluer. The LMC \& GES sample also has
the highest fraction of disc galaxies, 0.71, similar to the GES-o
sample.

\subsection{Satellite population}

\begin{figure}
    \centering
    \includegraphics[width=\columnwidth]{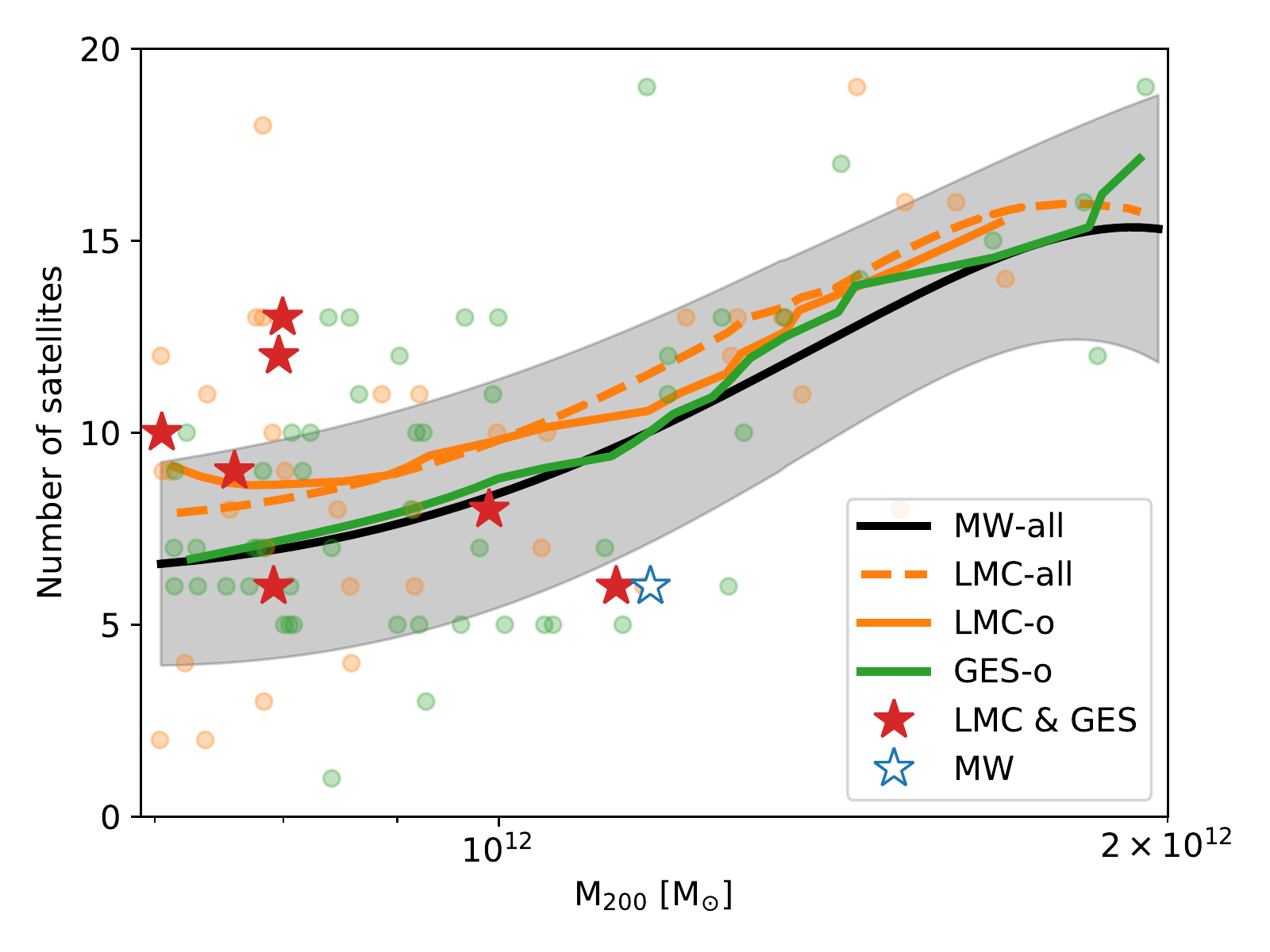}
    \caption{Number of satellites of stellar mass, $>1\times 10^6$
      M$_{\odot}$, within the radius R$_{200}$ for the host, as a
      function of host halo mass, M$_{200}$, for various samples of
      MW-mass galaxies. The colours, lines and shading are similar to
      those in Fig.~\ref{fig: star_halo hists}. The blue open star
      shows the number of Galactic satellites with stellar mass above
      $10^6$ M$_{\odot}$. R$_{200}$ and $M_{200}$ for the MW halo have
      been taken from \citet{callingham_mass_2019}. }
    \label{fig: satellites}
\end{figure}

In this subsection we examine the satellite population of our MW
samples and investigate whether the LMC \& GES sample is different in
this respect to the other samples. As shown in the simulations of the
APOSTLE project \citep{sawala_apostle_2016, fattahi_apostle_2016}, the
EAGLE model reproduces well the observed stellar mass function for
dwarfs galaxies, both those that are satellites orbiting in the MW and
those in the field around the Local Group. Several important
properties of dwarf galaxies, such as their sizes and star formation
histories, are also consistent with observations
\citep{campbell_knowing_2017,sawala_apostle_2016,bose2019,sales_black_2016,bose_little_2019,digby_star_2019}. 

The number of `luminous' satellites as a function of halo mass for our
various samples is shown in Fig.~\ref{fig: satellites}. Since EAGLE
does not resolve ultra-faint dwarfs, the satellites plotted are those
with stellar mass above $M_{*} > 1 \times 10^6 M_{\odot}$ that are
within the virial radius of the MW-mass haloes. The mass and number of
satellites above this stellar mass within the $M_{200}$ radius of our
Galaxy is shown as a blue open star symbol. Here, the halo mass of the
MW and its $R_{200}$ radius are taken from
\citet{callingham_mass_2019} and the stellar masses and Galactocentric
distances of the MW satellites from \citet{mcconnachie_observed_2012}.

Fig.~\ref{fig: satellites} shows the well-known correlation between
the number of satellites and the halo mass \citep{wang2012}.  The
LMC-all and LMC-o samples have slightly higher numbers of satellites
at fixed halo mass than the MW-all sample. This is probably because
the LMC-mass dwarfs would have brought in their own satellite cohorts
\citep{shao_evolution_2019}. The GES-o sample has a very similar trend to the
MW-all sample, possibly because these galaxies have a similar
late-time assembly history to the MW-all galaxies (Fig.~\ref{fig: m200
  vs z} right panel) when the bulk of the satellite population is
accreted \citep{fattahi_tale_2020}. At low halo masses, the infall of
the LMC-like satellite (in LMC-all, LMC-o and LMC \& GES samples) has
a much larger impact as it is fractionally much larger compared to the
host than for higher halo mass MW galaxies.  We also examined the
radial distribution of satellites and found no discernible differences
between the different categories of MW-like galaxies. However, we note
that we do not take ``orphan'' satellites into consideration, which
can make a substantial difference to the radial profiles
\citep{bose_little_2019}.

\section{Discussion and Conclusions}\label{sec: disc and conc}

We have used the EAGLE hydrodynamical simulation to assess the extent
to which the two known major accretion events experienced by our
Galaxy --the Gaia/Enceladus/Sausage (GES) merger at early times and
the more recent infall of the LMC-- have shaped the properties of our
galaxy. We have also tested whether these events make the MW stand out
amongst galaxies of similar mass.  

We identified 1078 MW-mass haloes in the mass range
$M_{200}=(0.7-2)\times10^{12}\Msun$ and subdivided this sample into
three: {\em (i)}~galaxies with an LMC-mass satellite within their
virial radius at $z=0$ (LMC-all sample); {\em (ii)}~galaxies with an
LMC as in {\em (i)}, but which did not experience a GES merger event
or accrete any other more massive dwarfs than that in the past 8~Gyrs
(LMC-o sample); {\em (iii)}~galaxies that experienced a merger between
$z=1-2$ similar to the GES but no further large mergers (GES-o
sample); and {\em (iv)}~galaxies that experienced both a GES merger
and the accretion of an LMC-mass satellite but no other significant
mergers in between these two (LMC \& GES sample). Only 7 galaxies fall
in the last category; their assembly histories bear the closest
resemblance to the assembly history of the MW, as far as it is
currently known. Our main conclusions are as follows:

\begin{itemize}

\item In agreement with earlier work, we find that the presence of an
  LMC-mass satellite orbiting a MW-mass galaxy is relatively uncommon:
  only 15.7\% of galaxies fall in our LMC-all category. However, once
  the constraint is imposed that there should have been no massive
  mergers in the past 8~Gyr, the frequency is reduced to 3.7\%
  (LMC-o). The number of MW-mass galaxies with a GES event between
  $z=1-2$ (GES-o) amounts to 5\% of the total sample.  These fractions
  are slightly higher than in other studies
  \citep[e.g.][]{cautun_aftermath_2019,bignone_gaia-enceladus_2019}
  because of different ways of identifying the LMCs and GES in
  simulations.

\item The assembly history of the MW is rare in the $\Lambda$CDM
  model: only 0.65\% of MW-mass EAGLE galaxies have both an early GES
  merger and a late LMC infall and no significant mergers in
  between. This sample most closely resembles the assembly history of
  the Milky Way.

\item The existence of an LMC or a GES event selects haloes towards the
  lower end of the mass range we considered. Requiring both events
  further lowers the halo mass.

\item At a fixed halo mass, galaxies with an LMC (with or without the
  restriction on past mergers) have lower than average stellar
  mass. This bias is partly a matter of definition since the stellar
  mass of the galaxy is measured within 30~kpc, well inside the
  position of a typical LMC, whereas the mass of the
  halo is measured out to $R_{200}$. 
\item Haloes destined to accrete an LMC have lower than average mass
  at early times but ``catch up'' at late times once the LMC has been
  accreted. The GES-o sample closely follows the accretion history of
  the MW-all sample until the merger with the GES around $z=2$, after
  which the increase in halo mass is much slower. Haloes in the LMC \&
  GES sample have much lower initial masses but they experience large
  increases in mass at the time of the merger with the GES and the
  later infall of the LMC. As a result, the formation redshift of the
  LMC \& GES haloes (defined as the time when 50\% of the final halo
  mass was in place) ends up being typical of haloes of that mass.

\item Although no constraints on the morphology, colour or star
  formation rate of the final galaxy were applied in the sample
  selection, the LMC \& GES galaxies all fall in the ``blue cloud'' in
  the colour-stellar mass diagram. This is a reflection of the long
  period without massive mergers required for this sample. Using a
  kinematical diagnostic of galaxy morphology (the fraction of the kinetic 
  energy invested in rotation), the LMC \& GES galaxies are
  predominantly disc galaxies: whereas 60\% of the MW-all sample are
  discs according to this definition, 70\% of the LMC \& GES sample
  are discs. 

\item Galaxies with an LMC have more satellites than galaxies without
  one, including those with a GES event. The excess is due to the
  additional satellite population brought into the MW halo by the
  LMC. 

\end{itemize}

Our simulations indicate that the build-up of both the stellar and
dark matter mass in the Milky Way were strongly influenced by the GES
merger and the accretion of the LMC. We therefore expect the
progenitors of the MW, perhaps accessible to observational study with
forthcoming telescopes such as JWST, to be atypical of galaxies of
similar mass today.

The main limitation of this work is the relatively small volume
sampled by the EAGLE simulation: there are only 7 galaxies in the
EAGLE volume that satisfy the main constraints we imposed for the
assembly history to resemble that of the MW, namely an early GES
merger, a late LMC infall and a quiescent phase in between. A larger
simulation is required to understand in more detail the atypical
properties of the assembly history of the MW.

\section*{Acknowledgements}
TE is supported by a Royal Society Research Grant, and AD by a Royal
Society University Research Fellowship. AF is supported by a
Marie-Curie COFUND/Durham Junior Research Fellowship (under grant
agreement no. 609412). CSF is supported by ERC Advanced Investigator
grant, DMIDAS [GA 786910]. We acknowledge support by the Science and
Technology Facilities Council (STFC) [grant numbers ST/T000244/1,
ST/P000541/1]. This work used the DiRAC Data Centric system at Durham
University, operated by the ICC on behalf of the STFC DiRAC HPC
Facility (www.dirac.ac.uk). This equipment was funded by BIS National
E-infrastructure capital grant ST/K00042X/1, STFC capital grant
ST/H008519/1, and STFC DiRAC Operations grant ST/K003267/1 and Durham
University. DiRAC is part of the National E-Infrastructure. This
research made use of the open source project \texttt{yt}
\citep{turk_yt_2010}.




\bibliographystyle{mnras}
\bibliography{references} 




\bsp	
\label{lastpage}
\end{document}